\documentclass[letterpaper]{article} 
\usepackage{aaai23}  
\usepackage{times}  
\usepackage{helvet}  
\usepackage{courier}  
\usepackage[hyphens]{url}  
\usepackage{graphicx} 
\usepackage{natbib}  
\usepackage{caption} 
\usepackage{algorithm}
\usepackage{algorithmic}

\usepackage{tikz}
\usetikzlibrary{shapes.symbols}
\usepackage{pgfplots}
\usepackage{booktabs}

\usepackage{newfloat}
\usepackage{listings}
\DeclareCaptionStyle{ruled}{labelfont=normalfont,labelsep=colon,strut=off} 
\floatstyle{ruled}
\newfloat{listing}{tb}{lst}{}
\floatname{listing}{Listing}
\title{This Sample seems to be good enough! Assessing Coverage and Temporal Reliability of Twitter's Academic API}
\author{
    J\"{u}rgen Pfeffer, \textsuperscript{\rm 1}
	Angelina Mooseder,	\textsuperscript{\rm 1}
	Jana Lasser, \textsuperscript{\rm 2}	
	Luca Hammer, \textsuperscript{\rm 3}\\	
	Oliver Stritzel, \textsuperscript{\rm 2}		
	David Garcia    \textsuperscript{\rm 2}
}
\affiliations{
    \textsuperscript{\rm 1} School of Social Science and Technology,
	Technical University of Munich, Germany\\
	\textsuperscript{\rm 2} Faculty of Informatics, Technical University of Vienna, Austria\\
	\textsuperscript{\rm 3} Department of Media Studies, University of Siegen, Germany\\

}

\usepackage{bibentry}

\begin{document}

\maketitle

\begin{abstract}
Because of its willingness to share data with academia and industry, Twitter has been the primary social media platform for scientific research as well as for consulting businesses and governments in the last decade. In recent years, a series of publications have studied and criticized Twitter's APIs and Twitter has partially adapted its existing data streams. The newest \emph{Twitter API for Academic Research} allows to "access Twitter's real-time and historical public data with additional features and functionality that support collecting more precise, complete, and unbiased datasets."\footnote{\url{https://developer.twitter.com/en/products/twitter-api/academic-research}}
The main new feature of this API is the possibility of accessing the full archive of all historic Tweets. In this article, we will take a closer look at the Academic API and will try to answer two questions. First, are the datasets collected with the Academic API complete? Secondly, since Twitter's Academic API delivers historic Tweets as represented on Twitter at the time of data collection, we need to understand how much data is lost over time due to Tweet and account removal from the platform. Our work shows evidence that Twitter's Academic API can indeed create (almost) complete samples of Twitter data based on a wide variety of search terms. We also provide evidence that Twitter's data endpoint v2 delivers better samples than the previously used endpoint v1.1. Furthermore, collecting Tweets with the Academic API at the time of studying a phenomenon rather than creating local archives of stored Tweets, allows for a straightforward way of following Twitter's developer agreement. Finally, we will also discuss technical artifacts and implications of the Academic API. We hope that our work can add another layer of understanding of Twitter data collections leading to more reliable studies of human behavior via social media data.
\end{abstract}

\section{Introduction}
In 2013, David Lazer opened his ICWSM \cite{icwsm2013} keynote address with the joke "Welcome to the $7^{th}$ International Conference on Twitter Analysis", referring to the fact that a majority of published papers at this conference were working with Twitter data. While a series of other social media platforms have attracted the attention of academic researchers in recent years, Twitter has remained the dominant data source for social media analysis of human behavior due to its data accessibility via a variety of application programming interfaces (API). Several review articles have tried to create an overview of the countless number of studies that have utilized Twitter data \cite{Yu2020,Williams2013,Zimmer2014,Jungherr2015,Sinnenberg2017,Karami2020,Pradyumn2018}.

For most of the data driven studies in recent years, two main data access strategies have been applied. Tweets were collected either with filtering queries (e.g. keywords, accounts, geo-locations) or as random 1\% or 10\% samples of all Tweets. Both data collection strategies relied on real-time data collection. With the Academic API, researchers are now able to perform full-archive searches, which has created a wide variety of new research possibilities.

In this article, we will build on the research that critically reflected Twitter's data sources and sampling algorithms. Similar to Zubiaga \cite{zubiaga_longitudinal_2018}, we will also re-collect historic data. We will then use the re-collected data and the error messages that the Twitter API creates for missing Tweets to assess how many Tweets are actually \emph{detectable} in the new data collection because they were neither deleted nor protected. We will also compare the collected data from the Academic API with data that was collected at the same time with the same search query via the costly Twitter Premium API. And we will perform controlled experiments to gain better insights into the sample creation process as well as into possible technical artifacts. Finally, we will discuss the implications of using the Academic API for studying Twitter activity. 

\textbf{Contributions.} The goal of this article is not to reverse engineer Twitter's hardware and software architecture. However, we do think that a certain level of technical insight into the data production process is necessary to assess the quality of the data source. The contributions of the article are:

\begin{itemize}
    \item We provide background information for researchers interested in using Twitter's Academic API.
    \item We show evidence that Twitter's Academic API can provide almost complete samples of the overall activity on Twitter related to specific keywords and time periods.
    \item We argue that Twitter's Academic API utilizing the new API v2 data endpoint delivers the best data samples from Twitter for most application scenarios.
    \item We study the proportion of Tweets that get lost over time when historic Twitter data is collected with full-archive searches.
    \item We will discuss technical artifacts and implications when collecting data with the Academic API.
    \item We offer a set of recommendations for researchers working with the Academic API in their research studies.
\end{itemize}

\section{Related Work}

\textbf{Twitter's Academic API.} Although the Twitter Academic API is relatively new, it has been used in multiple studies. Regarding the times in which the Twitter's Academic API has been released, it is not surprising that many of these studies cover Covid-19 related topics. Before the Academic API was released, researchers had to use services like Brandwatch to analyze exhaustive datasets around the pandemic \cite{Pellert2020,Metzler2021}. More recently, Aryal and Bhattarai \citeyearpar{aryal_sentiment_2021} used the API to collect Tweets by hashtags and geolocation and study sentiments towards Covid-19 vaccination across several Asian countries. Similarly, Butsos et al. \citeyearpar{bustos_twitter_2022} collected Tweets by hashtags to analyze the sentiments and emotions towards Covid-19 vaccination before and after the first launch of a Covid vaccination in the US. Marcec and Likic \citeyearpar{marcec_using_2021} compared the sentiments of English Tweets towards different Covid-19 vaccines over time and location, using the Academic API to collect Tweets by keywords, while Biswas et al. \citeyearpar{biswas_public_2022} collected Tweets by vaccine-related Arabic hashtags and analysed the sentiments of Arabic Tweets towards vaccination in general before and during Covid-19. 

A somewhat different data collection approach was applied by Muric et al. \citeyearpar{muric_covid-19_2021}. They first searched antivaccine related user accounts by employing a keyword search on Twitter's Streaming API. Afterwards they used the Academic API to collect historical Tweets of these accounts. Based on this data, they studied the topics and time distribution of anti-Covid-vaccination Tweets as well as the political leaning of accounts publishing such Tweets. 

Besides Covid-19 related studies, the Academic API has been applied in a great variety of research topics. Using searches based on hashtags or keywords, data was collected via the API to study topics of haemophilia-related Tweets and categorize users tweeting such content \cite{chen_digital_2022}; analyse the time distribution and user network of Tweets about an event of a false terrorist alarm at Oxford circus 2017 \cite{eriksson_krutrok_social_2022}; examine the topics, hashtags and mentions of polish Tweets about a Polish women's strike at two different timeframes \cite{paradowski_womens_2021}; investigate the topics of and user engagement in Tweets about women's day across three countries \cite{wallaschek_same_2022} and research the topics and time distribution of Tweets about the h-index \cite{thelwall_researchers_2021}. 

\textbf{Sampling quality and technological insights.}
Working with social media data is known to be challenged by different artifacts coming from data or applied methods \cite{Ruths2014}. In the past, researchers took a close look into the quality of Twitter data and possible errors, which might arise during data collection \cite{salvatore_social_2021}. They found that the choice of the API and the collection technique can heavily influence the accuracy, representativeness, and reproducibility of the dataset. 

A core challenge when using Twitter APIs is how to build a data sample which is representative of the overall activity on Twitter. Moerstaetter et al. \citeyearpar{Morstatter2013} compared datasets collected with the Streaming API and Firehose based on keywords, geolocation, and user account. They found that the representativeness of data collected with the Streaming API heavily depends on the coverage and type of analysis. For example, the results showed that the Streaming API performs worse than random samples of the Firehose datasets when coverage is low, but topical analysis results were similar between the Streaming API and Firehose datasets when coverage was high. 
Likewise, Wang et al. \citeyearpar{wang_yazhe_and_callan_jamie_and_zheng_baihua_should_2015} used the Twitter Streaming API with Spritzer and Gardenhose access to collect Tweets of specific user accounts and compare them to a dataset of all Tweets of these users collected with the REST API. Their results indicated that data collected with the Streaming API provides enough information to analyse general or content statistics, but only covers small proportions of user interaction, such as mentions or Retweets.
Furthermore, Kim et al. \citeyearpar{kim_story_2020} compared datasets collected with the filtered Streaming API, Search API and GNIP historic Powertrack (Firehose). They found that the volume, content and accounts of Tweets within these datasets varied substantially and that the use of the historic API may result in an underestimation of the amount of marketing and bot-generated Tweets and accounts. Also, Alkulaib et al. \citeyearpar{alkulaib_collect_2020} compared datasets of event-related Tweets collected with the Streaming API, Search API and Firehose API and found only a small overlap between these datasets. They suggested that a combination of multiple APIs could results in a more representative dataset. This strategy was also recommended by Timoneda \citeyearpar{timoneda_joan_where_2018}, who compared datasets collected with the Streaming and the Search API based on keywords and found that while 20-30\% of Tweets did not appear in the Search API, also on average between 2 and 5\% of Tweets were lost during data collection with the Streaming API. Similarly, different collection tools can influence the resulting data sample. For example, Weber et al. \citeyearpar{weber_exploring_2021} compared datasets collected with Twarc, RAPID and Tweepy, all based on the filtered Streaming Twitter API, and found considerable differences between these datasets, which significantly altered the results of subsequent network analyses. Wu et al. \citeyearpar{Wu2020} studied Twitter sampling effects across different timescales and different subjects by analyzing the 
rate limit messages that indicate the cumulative number of missing tweets for an API connection.

One major problem is that there has been little information from Twitter about how data is sampled by the APIs and that biases might be introduced by sampling mechanisms \cite{chen_twitter_2021}. Kergel et al. \citeyearpar{Kergl2014} were instrumental in revealing the sampling procedure of the $1\%$ and the $10\%$ Sample API. They were able to show meta-data extracted from the Tweet ID can tell us whether a Tweet is in one of Twitter's Sample streams or not. Another question is, whether data collected with Twitter's APIs are reproducible. Zubiaga \citeyearpar{zubiaga_longitudinal_2018} collected 30 datasets in 2012-2016 using the Twitter Streaming API. In 2016, he recollected the data, using the API's statuses/lookup method based on Tweet id, to analyze how much of the data can be reproduced. He found a substantial loss of Tweets and user accounts, with some datasets having more than 30\% of data that could not be reproduced, as well as a change in aggregated user metadata. Nevertheless, he showed that the datasets were still representative in regard to textual content.

\section{Twitter's data streams}

\textbf{Twitter's data APIs.} Based on underlying functionality, Twitter offers different types of APIs for accessing data. First, the Filter API collects Tweets which include user defined search terms (e.g. keywords, accounts). Second, the Sample API provides a sample (1\% or 10\%) of \emph{all} Tweets. It is important to understand that the Sample and the Filter API (together often referred to as the Streaming API) offer real-time data collection, i.e. the Tweets of our data collection are coming as a \emph{stream} through the API immediately after they have been sent. In contrast, the REST and the Search API can be used to collect historic Tweets and other non-real-time information. The REST (Representational State Transfer) API can be used to collect information about accounts and their followers and followees or to send Tweets. Finally, Search APIs (including the Academic API) can search for historic Tweets (either only some days back or all the way back to the first Tweets in 2006). Several of these APIs offer different access levels, e.g., Standard, Premium, Academic, Enterprise, with elevated access restricted to specific user groups or paying clients. 

\textbf{Data endpoints.} The Academic API was introduced as part of a new data endpoint, that Twitter calls its API v2. While there are several minor differences, the major difference when collecting data is that with the old API v1.1, a Tweet in JSON format always arrived as a single complete object, i.e. many of the Tweet's fields are delivered automatically. The endpoint API v2 just delivers the id and the text field of a Tweet automatically; all other fields need to be requested via call parameters. Furthermore, account information of the sender and other meta data comes in a separate JSON object. Another obvious difference between these two data endpoints is that users traditionally needed an authentication key consisting of four parts with the v1.1 endpoints while a single longer \emph{bearer token} key is needed for the v2 endpoint.

\textbf{Implications when collecting historic Tweets.} In recent years, most researchers had collected Twitter data with the real-time sample and filter APIs. By utilizing the Academic API, the data collections strategy changes, which comes with certain implications that we will discuss in this paper. First and foremost, when collecting historic Tweets with the Academic API, we will get a representation of Twitter as of the collection time and \emph{not} the time when the Tweets were created. Simply speaking, the Academic API can access the same information that you can access at Twitter's website (e.g. via the search feature). This also implies that Tweets that have been deleted by users or when an account is removed (e.g. the suspended account @realdonaldtrump) from Twitter, will not be part of a dataset collected with the Academic API.

Another very important aspect comes from the fact that Twitter seems to not store information about the sender of a Tweet (except the account ID) with the Tweet. Consequently, when you collect historic Tweets and also include user fields, these fields are requested from the user database at the time when the Tweet is collected and not at the time when the Tweet was created. This means that the user profile information and even the user name (but not the user ID) could have changed in the meantime.

\section{Experiments to assess data quality}

In this section we present the setup and results of several experiments that were designed to assess the sampling quality of the Academic API. We do this by studying the two key aspects of this API, the full-archive search and the data endpoint v2. Because we do not have direct access to Twitter's database of Tweets, we need to design different experiments to get different insights into the software and hardware black box. However, we start with a controlled experiment for which we actually know the ground truth.

\subsection{Controlled experiment}
We performed a controlled experiment to identify possible obvious issues with the data streams. With a set of ten Twitter accounts, we sent $3,400$ Tweets within $2.5$ hours---one Tweet every $~2.5$ seconds---with a single random $42-$character string. We had setup two filter API scripts listening for Tweets with the same string, one each utilizing the v1.1 and the v2 data endpoint. After sending the Tweets, we also collected the Tweets from the previous hours with the Academic API, again with a query searching for the specific string. We repeated this search with the Academic API $24$ hours later. For all four data collection approaches, we can report that $100\%$ of all Tweets were collected. In other words, if your search queries include a smaller number of Tweets loosely distributed over time, there is a very good chance that you will receive a very high coverage, most likely with all of Twitter's APIs and endpoints. 

\subsection{High volume topics}
\begin{figure}[h]
\centering
\includegraphics[width=0.65\linewidth]{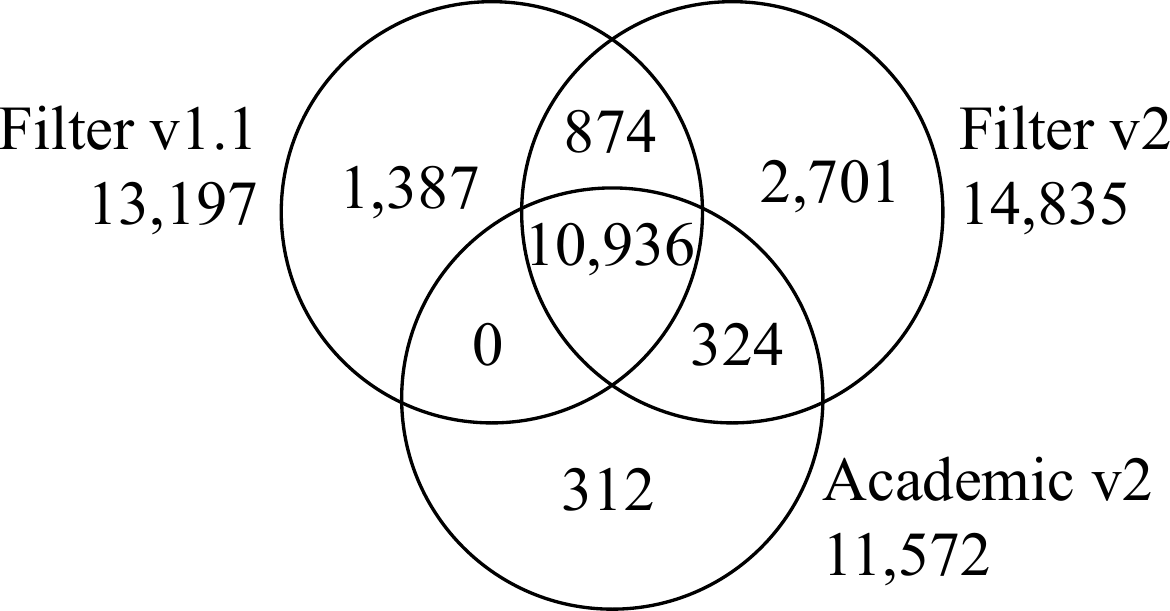}
\caption{Symbolic Venn diagram of the overlap of three different data sources in a high volume experiment. }
\label{fig:venn} 
\end{figure}

In a subsequent experiment, we were interested in the performance of these data collection approaches when dealing with a topic with a high volume of Tweets. Since our study was conducted during the Russian war in Ukraine, we searched for the term "Ukraine" for exactly five minutes via the v1.1 and v2 data endpoints in parallel on the same computer. We are aware that Twitter tries to prevent a single query collecting a very large number of Tweets and, consequently, inserts rate limits into the Filter API. Independently from rate limits, our primary interest is in coverage, i.e. more is better. Right after the period of live data collection, we collected "Ukraine" Tweets with the Academic API for the exact same five minutes. Surprisingly, the results were very different with this setup (see Figure \ref{fig:venn}).

While the numbers between the live collection and the historic collection with the Academic API cannot be compared directly in terms of coverage, since Tweets can be protected (made \emph{invisible}) and unprotected or deleted, some of the numbers in figure \ref{fig:venn} are surprising. The dynamics as well as different rate limits result in the fact that only $72.8\%$ of Tweets collected with these three approaches occur in all datasets. The difference between the two Filter APIs indicates that the two data endpoints in fact produce different datasets and that the endpoint v2 creates a larger dataset. The relatively high number of Tweets that only appear in the Filter APIs are strong indicators that a significant Twitter volume gets removed from the platform quickly, either because users delete or hide their Tweets or Twitter removes Tweets because of violations of community norms (e.g. toxic language). Another very interesting number in this Venn diagram is the number $0$ on the intersection of \emph{Academic v2} and \emph{Filter v1.1}. This means that there was not a single Tweet that was picked up by the \emph{Filter v1.1} that was still available for the later data collection with the \emph{Academic v2} but had not already been picked up by the \emph{Filter v2}. More proof that the two data endpoints are technically different and another indicator that the endpoint v2 is superior because of the higher number of delivered Tweets.

\textbf{Time delay.} We can re-construct the arrival time in milliseconds of a Tweet at one of Twitter's servers \cite{Pfeffer2018Tampering,Kergl2014} from the Tweet ID. Having around $40-50$ Tweets per second in our datasets, we took a closer look at these timestamps and, despite possible time differences between Twitter's servers and our experimental computers, we made two observations:

\begin{enumerate}
    \item With the Filter v1.1 API, a Tweet took on average $4.4$ seconds [$4.3-5.7$] to get to our computer, with $90\%$ arriving within $4.7$ seconds. In contrast, with the Filter v2 it took on average $10.2$ seconds [$9.4-19.7$] with almost $90\%$ arriving within $10 +/- 0.5$ seconds so that we can assume that $10$ seconds is Twitter's new intended time delay. 
    
    \item Tweets do not come in perfect temporal order at either data endpoint. However, with the Filter v1.1 API, less than $1\%$ of Tweets arrived more than $0.2$ seconds after a Tweet that they were supposed to precede, with no single Tweet changing the order for more than $1$ second. With the Filter v2 API these numbers change dramatically. More than $40\%$ of Tweets arrive more than $0.2$ seconds after a Tweet that they were supposed to precede, with a maximum of almost $9$ seconds. 
\end{enumerate}

The first point is consistent with the minimum time period that needs to pass in order to search for historic Tweets with the Academic API. The second point seems to be another indication that the v2 data endpoint is in fact a different system architecture. We assume that these two aspects combined create a situation in which, for the Filter v2, more Tweets are allowed to get onto the API server before they are picked up by the \emph{real-time} API requests.

We will not speculate here any further but, supported by more experimental runs with different search terms, we can conclude that in general, the Filter v2 API will return more Tweets.

\subsection{Searching for historic Tweets}
Another test that we performed in order to assess the characteristics of the Academic API is to re-collect data that was collected $8-11$ years ago with the $10\%$ Decahose Sample API when studying online firestorms \cite{Lamba2015}. What we expected is that a) we would get much more data with the Academic API today and b) that there should be no Tweet in the old dataset that is not shown today and that has not been deleted or protected. Answering the first point is straightforward. When the dataset was collected in real-time with the then-used Twitter Decahose Sample API, $\mathcal{D} = 165,267$ Tweets were collected for the 20 online firestorms. When collecting Tweets again on March 6, 2022, with the same hashtag/user account queries for the same time periods in $2011-2014$ using the Academic API and utilizing Twitter's data API v2, we were able to collect $\mathcal{A}_1 = 799,033$ Tweets.  

\begin{figure}[t]
\centering
\footnotesize
\begin{tikzpicture}
\draw (2,2) rectangle (4,3.8) node[pos=.5, align=center] {165,267 \\ Tweets};
\draw (4.4,2) rectangle (6,2.9) node[pos=.5, align=center] {74,359 \\ Available};
\draw (4.4,2.9) rectangle (6,3.8) node[pos=.5, align=center] {90,980 \\ Missing};
\draw (8.2,2) rectangle (9.5,7) node[pos=.5, align=center] {799,033 \\ Tweets };
\draw[->] (4,2.9) -- (4.4, 2.9);
\draw[->] (6,2.3) -- (8.2, 2.3);
\draw[->] (3,5) -- (3,3.8);
\draw[->] (4,6) -- (8.2, 6);
\node[cloud, draw, fill = gray!10, minimum height= 3cm, minimum width= 4cm] at (3.1,6) {} ;
\node[text width=3.5cm, align=center] at (3,6) {\textbf{ $\approx{}$1,652,670 Tweets \\ Created 2011-2014}};
\node[text width=3cm, font=\scriptsize] at (7.7,3.4) 
    {66,278 Not found \\ 24,209 Authoriz. \\ 421 Tweets found};
\node[text width=3cm, font=\scriptsize] at (8,2.1) {45,0\% };
\node[text width=3cm, font=\scriptsize] at (7.5,5.8) {48,3\% };
\node[text width=3cm, font=\scriptsize ] at (3.2,1.6) 
    {10\% Sample Decahose v1 \\ Collected Live 2011-2014};
\node[text width=3cm, font=\scriptsize , align=center] at (8.7,1.6) 
    {Academic API v2 \\ Collected March 6,2022};
\end{tikzpicture}
\caption{Overview of data samples for comparing Academic API data with historic Sample API data.}
\label{fig:deca} 
\end{figure}
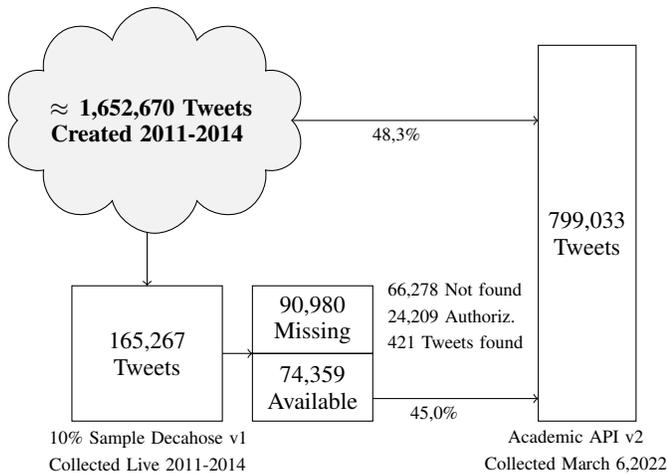

It is important to understand the above-described differences between these two different data collection methods. We can assume that $\mathcal{D}$ was about $10\%$ of ${\sim}1.65$ million Tweets (see Fig. \ref{fig:deca}). Comparing this number with $\mathcal{A}_1$ implies that slightly more than half of all Tweets from $\mathcal{D}$ cannot be found on Twitter anymore ${\sim}10$ years later. However, this observation does not answer our question as to whether we were able to collect all Tweets in 2022 that are still on the platform from 2011-2014. Missing Tweets can be due to different reasons. Tweets can be deleted, they can be protected (i.e. not publicly visible), or they could be missing in our sample while still being on Twitter. To distinguish between these three cases, we first compared the two lists of Tweet IDs and identified the 90,908 Tweet IDs that were in the old dataset but were missing in the new dataset. In a second step, we tried to collect the Tweets of these missing Tweet IDs and take advantage of the fact that Twitter delivers different error messages for different reasons for Tweets being untraceable. For $72.9\%$ (66,278 Tweets) we received a "not found error" indicating that a Tweet is no longer on the platform. For $26.6\%$ (24,209 Tweets) we received an "authorization error" pointing to the situation that these Tweets being protected by their users at the time of data collection. Only 421 Tweets ($0.5\%$) from $\mathcal{D}$ that were not in $\mathcal{A}_1$ could be found. These are Tweets that potentially were missed in the Academic API search. We repeated the data collection describe above about 5 hours after the first collection. We performed the same search queries and received $\mathcal{A}_2 = 799,027$ Tweets, i.e. six Tweets less. It is highly likely that six out of the original almost 800k Tweets were deleted within five hours, e.g. via the removal of a user account. Because of the small proportion of missing Tweets, we did not follow this path any longer and conclude that the Academic API can successfully collect all (or at least almost all) Tweets for a search term and time period that are still on the platform at the time of data collection. This example also shows how, in general, historic Sample API data can be used to estimate the number of missing Tweets when collecting data with the Academic API.

\subsection{Verifying results with Twitter's Premium API}

Another approach of assessing the quality of the dataset collected with the Academic API is by comparing it with a different data source that is supposed to extract the same dataset from Twitter. The Twitter Premium API is a paid-for service for enterprises that gives users access to the "the full history of Twitter data". We were able to access the Premium API and collected the same data as described above, at the same time of collecting $\mathcal{A}_1$. Interestingly, the Premium API delivered only $\mathcal{P} = 774,485$ Tweets. When comparing the Tweet IDs of the two data sources, 24,978 Tweets from $\mathcal{A}_1$ are missing in $\mathcal{P}$, while only 430 Tweets from $\mathcal{P}$ are missing in $\mathcal{D}_1$. At this point, we can only hypothesize that the API endpoint v1.1 has a different approach for searching in Twitter's archival data that leads to different results. Of course, there is a possibility that the endpoint v2 also delivers Tweets that should not be part of the queried result, but most likely the endpoint v1.1 misses Tweets with certain properties due storage or search characteristics---further research can look more closely into these differences. At this point, we can report that the Academic API delivers more Tweets in response to a historic search than the costly Premium API utilizing endpoint v1.1.

\section{Tweet data decay}\label{sec:decay}

A major feature of the Academic API is that most researchers collect their data a certain time after an event and not in real-time. In the previous section we could see that about half of all Tweets are lost after $10$ years. In this section, we will first take a closer look at the first hours and days after Tweets were sent. Secondly, we will try to understand the long-term Tweet decay that follows a simple exponential decay function. 

\begin{figure}[t]
\centering
\footnotesize
\begin{tikzpicture}
\node[font=\tiny, align =center] at (0,2) {Day 1-7};
\draw (0.5,1) -- (2.5,1);
\draw (4.5,1) -- (6.5,1);
\draw [densely dotted] (0.5,1) -- (8.1,1);

\draw (0.7,1) -- (0.7,1.2) node[pos=4.3, rotate=90, font=\tiny] {06:42:42:000};
\node[text width=3cm, align=center, font=\tiny] at (1,1) {data \\collection};
\draw[->] (1,1.3) -- (1,2.5);

\draw (4.7,1) -- (4.7,1.2) node[pos=4.3, rotate=90, font=\tiny] {07:42:42:000};
\node[text width=3cm, align=center, font=\tiny] at (5,1) {data \\collection};
\draw[->] (5,1.3) -- (5,2.5);

\draw[->, line width=1mm] (7.6,1.5) -- (8.1,1.5);
\node[text width=3cm, align=center, font=\tiny] at (7.8,1.9) {24 hours};

\node[font=\tiny, align =center] at (0,4) {Day 0 \\ 3/9/2022};
\draw (0.5,3.1) -- (1.5,3.1);
\draw (2.5,3.1) -- (3.5,3.1);
\draw (4.5,3.1) -- (5.5,3.1);
\draw (6.5,3.1) -- (7.5,3.1);
\draw [densely dotted] (0.5,3.1) -- (8.1,3.1);

\draw (0.7,3.1) -- (0.7,3.3) node[pos=4.3, rotate=90, font=\tiny] {06:42:42:000};
\draw (1.3,3.1) -- (1.3,3.3) node[pos=4.3, rotate=90, font=\tiny] {06:42:42:999};
\draw (2.8,3.1) -- (2.8,3.3) node[pos=4.3, rotate=90, font=\tiny] {06:42:53:000};
\node[text width=3cm, align=center, font=\tiny] at (1,3) {\textbf{1 sec \\observation \\ period}};
\node[text width=3cm, align=center, font=\tiny] at (3.1,3.1) {data \\ collection};

\draw (4.7,3.1) -- (4.7,3.3) node[pos=4.3, rotate=90, font=\tiny] {07:42:42:000};
\draw (5.3,3.1) -- (5.3,3.3) node[pos=4.3, rotate=90, font=\tiny] {07:42:42:999};
\node[text width=3cm, align=center, font=\tiny] at (5,3) {\textbf{1 sec \\observation \\ period}};

\draw (6.7,3.1) -- (6.7,3.3) node[pos=4.3, rotate=90, font=\tiny] {07:42:53:000};
\node[text width=3cm, align=center, font=\tiny] at (7,3.1) {data \\ collection};
\draw[->, line width=1mm] (7.6,3.5) -- (8.1,3.5);
\node[text width=3cm, align=center, font=\tiny] at (7.8,3.9) {24 hours};

\draw[bend left=50][->] (3.1,2.8) to (1.1,2.5);
\draw[bend left=50][->] (7,2.8) to (5.1,2.5);

\end{tikzpicture}
\caption{Experimental setup for testing Tweet decay over a ten day time period. }
\label{fig:setup} 
\end{figure}
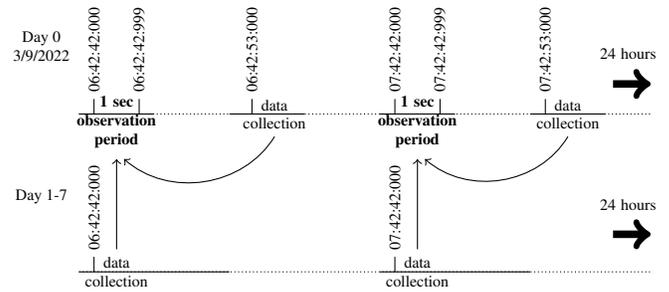

Our experimental setup for collecting data is depicted in Fig. \ref{fig:setup}. Our search term was constructed so that we were able (to the best of our knowledge) to collect all Tweets sent within a specific second. The first time-period for data collection was March 9, 2022\footnote{The initial dataset was collected March 9 \& 10, 2022---notably including the Russian war in Ukraine and Twitter's service blockage in Russia. Since we are not analyzing the content of the Tweets, these and other events should not impact the following analysis.} at 6:42:42am UTC collecting Tweets within the [$0-999$] millisecond range of this second on Twitter. To not bias against individual time zones, we collected the same second of Twitter data every hour for $24$ hours. The collection happened $10$ seconds after each time window---due to API restrictions the time window of data collection must end a minimum of 10 seconds prior to the request time---and took 7--15 seconds each. We call this the $day_0$ data. On average, we collected 4,526 Tweets per 1 second of observation period over 24 hours within a range of [3,374--5,953].

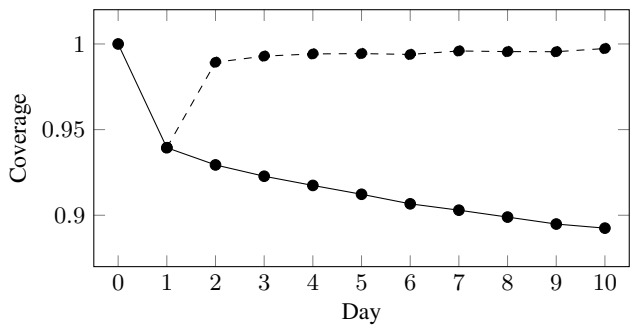
\begin{figure}[t]
\centering
\footnotesize
\begin{tikzpicture}
\begin{axis}[
	xlabel=Day,
	ylabel=Coverage,
	width=8.7cm,height=5cm,
	xmax=10.5,ymax=1.02,ymin=0.87,xmin=-0.5,
	ytick={0.9,0.95,1.0},
	xtick={0,1,2,3,4,5,6,7,8,9,10},
    y label style={yshift=-.8em}, x label style={yshift=.5em}]
\addplot [dashed,mark=*]
table[x index=0, y index=1] {
1 0.93943
2 0.98930
3 0.99290
4 0.99418
5 0.99438
6 0.99388
7 0.99591
8 0.99554
9 0.99548
10 0.99733
};
\addplot [black,mark=*]
table[x index=0, y index=1] {
0 1
1 0.93943
2 0.92938
3 0.92278
4 0.91741
5 0.91225
6 0.90667
7 0.90296
8 0.89893
9 0.89487
10 0.89248
};
\end{axis}
\end{tikzpicture}
\caption{Coverage over time. Proportion of Tweets from $day_0$ data in $day_n$ data (solid line). Proportion of Tweets from $day_{n-1}$ data in $day_n$ data (dashed line).}
\label{fig:coverage} 
\end{figure}

On the following day, we re-collected the data from $day_0$ with the identical search query and time periods exactly $24$ hours after the Tweets were created. This initial re-collection of $day_0$ data was then repeated for ten days to assess the Tweet decay ratio per day. Fig. \ref{fig:coverage} plots the results of our experimental data collection. The major drop in Tweet availability ($-6.1\%$) happens within the first $24$ hours. This is a very intuitive result. Twitter users delete Tweets because of typos or other reasons; Tweets get flagged as toxic and removed from Twitter; etc. Starting with $day_2$, the daily decay becomes smaller and smaller and  after ten days, about $10.8\%$ of Tweets are not available anymore.

\subsection{Short-term vs. long-term decay}
To get a better estimate of the  short-term (hours) and the long-term   (months) proportions of Tweets that are lost for an Academic API search, we utilized archived IDs from Twitter's 1\% data stream at \emph{archive.com} and another repository \cite{Lasser1p2022} over a time span of 5 years. For every data point of the following analysis, we collected all Tweet IDs for the given time span (month, day or hour) and drew five random samples of $100k$ Tweet IDs each. Each sample is then re-hydrated using the twarc2\footnote{\url{https://twarc-project.readthedocs.io/en/latest/twarc2_en_us/}} hydrate functionality. The number of available tweets ("found") and unavailable tweets ("not found" and "authorization error") are counted. The data points in the following figures are averages over the five samples for every point in time. The standard deviation for the points is so small ($<$0.01\%) that error bars for these data points are not visible. For the shorter time scales, we re-hydrated the data with a number of API keys in parallel, to reduce artifacts resulting from longer data collection time periods. 

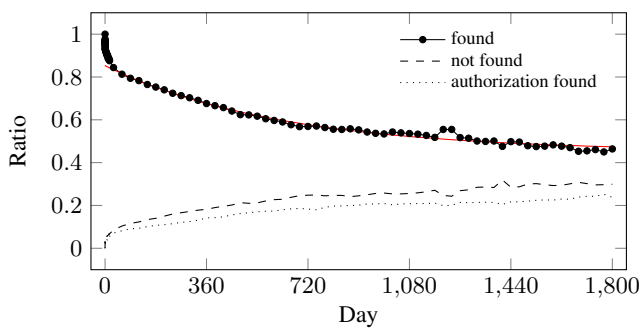
\begin{figure}[b]
\centering
\footnotesize
\pgfplotsset{scaled y ticks=false}
\begin{tikzpicture}
\begin{axis}[
	xlabel=Day,
	ylabel=Ratio,
	width=8.7cm,height=5cm,	xmax=1850,ymax=1.1,ymin=-0.1,xmin=-50,
	ytick={0,0.2,0.4,0.6,0.8,1.0},
	xtick={0,360,720,1080,1440,1800},
    y label style={yshift=-.8em}, x label style={yshift=.5em},
        legend style={cells={anchor=west},
        draw=none, fill=none, font=\scriptsize,
        legend pos=north east, row sep=-2pt}
    ]
\addplot [solid, mark=*, mark options={scale=0.6} ]
table[x index=0, y index=1] {
0.000 1.000
0.042 0.977
0.083 0.971
0.125 0.969
0.167 0.966
0.208 0.966
0.250 0.962
0.292 0.961
0.333 0.962
0.375 0.960
0.417 0.957
0.458 0.953
0.500 0.952
0.542 0.953
0.583 0.953
0.625 0.954
0.667 0.951
0.708 0.948
0.750 0.944
0.792 0.939
0.833 0.931
0.875 0.938
0.917 0.934
0.958 0.933
1.000 0.928
1.000 0.927
2.000 0.920
3.000 0.919
4.000 0.916
5.000 0.910
6.000 0.905
7.000 0.903
8.000 0.902
9.000 0.902
10.000 0.894
11.000 0.887
12.000 0.885
13.000 0.888
14.000 0.882
15.000 0.880
16.000 0.876
30.000 0.844
60.000 0.813
90.000 0.794
120.000 0.783
150.000 0.765
180.000 0.753
210.000 0.740
240.000 0.724
270.000 0.713
300.000 0.703
330.000 0.691
360.000 0.676
390.000 0.666
420.000 0.657
450.000 0.641
480.000 0.624
510.000 0.623
540.000 0.617
570.000 0.605
600.000 0.597
630.000 0.590
660.000 0.577
690.000 0.568
720.000 0.569
750.000 0.571
780.000 0.564
810.000 0.556
840.000 0.555
870.000 0.558
900.000 0.553
930.000 0.543
960.000 0.537
990.000 0.534
1020.000 0.543
1050.000 0.539
1080.000 0.536
1110.000 0.533
1140.000 0.527
1170.000 0.518
1200.000 0.555
1230.000 0.555
1260.000 0.517
1290.000 0.514
1320.000 0.501
1350.000 0.499
1380.000 0.501
1410.000 0.476
1440.000 0.498
1470.000 0.496
1500.000 0.479
1530.000 0.475
1560.000 0.477
1590.000 0.483
1620.000 0.477
1650.000 0.470
1680.000 0.453
1710.000 0.455
1740.000 0.461
1770.000 0.450
1800.000 0.464
};
\addplot [dashed]
table[x index=0, y index=1] {
0.000 0.000
0.042 0.020
0.083 0.024
0.125 0.026
0.167 0.028
0.208 0.027
0.250 0.027
0.292 0.028
0.333 0.027
0.375 0.028
0.417 0.029
0.458 0.030
0.500 0.030
0.542 0.031
0.583 0.031
0.625 0.032
0.667 0.033
0.708 0.034
0.750 0.039
0.792 0.042
0.833 0.042
0.875 0.036
0.917 0.037
0.958 0.040
1.000 0.042
1.000 0.041
2.000 0.047
3.000 0.047
4.000 0.051
5.000 0.053
6.000 0.056
7.000 0.057
8.000 0.059
9.000 0.058
10.000 0.060
11.000 0.062
12.000 0.063
13.000 0.064
14.000 0.067
15.000 0.066
16.000 0.068
30.000 0.085
60.000 0.104
90.000 0.116
120.000 0.123
150.000 0.134
180.000 0.140
210.000 0.150
240.000 0.160
270.000 0.166
300.000 0.172
330.000 0.178
360.000 0.182
390.000 0.190
420.000 0.196
450.000 0.202
480.000 0.213
510.000 0.210
540.000 0.212
570.000 0.221
600.000 0.227
630.000 0.229
660.000 0.238
690.000 0.246
720.000 0.248
750.000 0.249
780.000 0.245
810.000 0.247
840.000 0.247
870.000 0.241
900.000 0.245
930.000 0.250
960.000 0.256
990.000 0.258
1020.000 0.254
1050.000 0.254
1080.000 0.256
1110.000 0.259
1140.000 0.264
1170.000 0.271
1200.000 0.247
1230.000 0.243
1260.000 0.269
1290.000 0.273
1320.000 0.285
1350.000 0.286
1380.000 0.284
1410.000 0.319
1440.000 0.285
1470.000 0.287
1500.000 0.300
1530.000 0.302
1560.000 0.297
1590.000 0.293
1620.000 0.294
1650.000 0.301
1680.000 0.308
1710.000 0.302
1740.000 0.296
1770.000 0.297
1800.000 0.299
};
\addplot [dotted]
table[x index=0, y index=1] {
0.000 0.000
0.042 0.003
0.083 0.005
0.125 0.005
0.167 0.006
0.208 0.007
0.250 0.011
0.292 0.011
0.333 0.011
0.375 0.012
0.417 0.014
0.458 0.016
0.500 0.018
0.542 0.017
0.583 0.016
0.625 0.014
0.667 0.015
0.708 0.018
0.750 0.017
0.792 0.019
0.833 0.027
0.875 0.026
0.917 0.029
0.958 0.027
1.000 0.029
1.000 0.032
2.000 0.033
3.000 0.034
4.000 0.033
5.000 0.036
6.000 0.039
7.000 0.039
8.000 0.039
9.000 0.040
10.000 0.045
11.000 0.051
12.000 0.052
13.000 0.048
14.000 0.051
15.000 0.054
16.000 0.056
30.000 0.071
60.000 0.082
90.000 0.091
120.000 0.093
150.000 0.101
180.000 0.107
210.000 0.110
240.000 0.115
270.000 0.121
300.000 0.125
330.000 0.131
360.000 0.142
390.000 0.143
420.000 0.147
450.000 0.156
480.000 0.162
510.000 0.166
540.000 0.170
570.000 0.173
600.000 0.174
630.000 0.180
660.000 0.184
690.000 0.186
720.000 0.183
750.000 0.180
780.000 0.191
810.000 0.197
840.000 0.198
870.000 0.200
900.000 0.201
930.000 0.207
960.000 0.207
990.000 0.209
1020.000 0.203
1050.000 0.207
1080.000 0.208
1110.000 0.209
1140.000 0.209
1170.000 0.211
1200.000 0.199
1230.000 0.201
1260.000 0.214
1290.000 0.213
1320.000 0.214
1350.000 0.215
1380.000 0.214
1410.000 0.205
1440.000 0.217
1470.000 0.217
1500.000 0.220
1530.000 0.224
1560.000 0.226
1590.000 0.225
1620.000 0.230
1650.000 0.229
1680.000 0.239
1710.000 0.243
1740.000 0.243
1770.000 0.252
1800.000 0.238
};
\addplot [red]
table[x index=0, y index=1] {
0.000 0.853
0.042 0.853
0.083 0.853
0.125 0.853
0.167 0.853
0.208 0.853
0.250 0.853
0.292 0.853
0.333 0.853
0.375 0.853
0.417 0.853
0.458 0.853
0.500 0.852
0.542 0.852
0.583 0.852
0.625 0.852
0.667 0.852
0.708 0.852
0.750 0.852
0.792 0.852
0.833 0.852
0.875 0.852
0.917 0.852
0.958 0.852
1.000 0.852
1.000 0.852
2.000 0.852
3.000 0.851
4.000 0.850
5.000 0.850
6.000 0.849
7.000 0.848
8.000 0.848
9.000 0.847
10.000 0.846
11.000 0.846
12.000 0.845
13.000 0.844
14.000 0.844
15.000 0.843
16.000 0.843
30.000 0.834
60.000 0.816
90.000 0.799
120.000 0.782
150.000 0.767
180.000 0.752
210.000 0.738
240.000 0.724
270.000 0.711
300.000 0.699
330.000 0.687
360.000 0.676
390.000 0.666
420.000 0.655
450.000 0.646
480.000 0.637
510.000 0.628
540.000 0.620
570.000 0.612
600.000 0.604
630.000 0.597
660.000 0.590
690.000 0.584
720.000 0.577
750.000 0.571
780.000 0.566
810.000 0.560
840.000 0.555
870.000 0.550
900.000 0.546
930.000 0.541
960.000 0.537
990.000 0.533
1020.000 0.529
1050.000 0.526
1080.000 0.522
1110.000 0.519
1140.000 0.516
1170.000 0.513
1200.000 0.510
1230.000 0.507
1260.000 0.505
1290.000 0.502
1320.000 0.500
1350.000 0.497
1380.000 0.495
1410.000 0.493
1440.000 0.491
1470.000 0.489
1500.000 0.488
1530.000 0.486
1560.000 0.484
1590.000 0.483
1620.000 0.481
1650.000 0.480
1680.000 0.479
1710.000 0.477
1740.000 0.476
1770.000 0.475
1800.000 0.474
};
\addlegendentry{found}
\addlegendentry{not found}
\addlegendentry{authorization found}
\end{axis}
\end{tikzpicture}
\caption{Proportion of Tweets after $n$ days that are found and not found.}
\label{fig:historic_vanished}
\end{figure}

Figure \ref{fig:historic_vanished} shows the results of this analysis by charting the proportion of Tweets that could be found $n$ days after being sent, as well as the proportion of Tweets with \emph{authorization error} and \emph{not found} errors.  

Our subsequent intention with these numbers was to fit a simple exponential decay function of available Tweets, so that the coverage of historic data collections can be estimated in future research. Estimating the proportion of available Tweets can be an important indicator for assessing the fidelity of applied metrics. We have seen in Figure \ref{fig:coverage}, that the daily decay slows down dramatically after the first day and that the proportion of Tweets that remain available from one day to the next reaches $99.7\%$ on day $10$. This proportion continues to grow. However, after one month $84.4\%$ of Tweets could be found and this number is reduced on average by about $2\%$ per month for the first $1.5$ years. The rate of decay is then reduced to about $0.6\%$ per month. 

More precisely, the decay can be fit very well with an exponential decay function if we only consider data points of Tweets $1$ month or older. For this month time scale, we employ a function of the form

\begin{equation}
    f(x) = A \cdot exp(K \cdot x) + C.
    \label{eq:model}
\end{equation}

With $x$ being the age of Tweets in \emph{days}, we performed a non-linear least squares fit weighted by the standard deviations for each data point to estimate the remaining variables. Since for our data, the constant $C$ in eq. \ref{eq:model} can be interpreted as a lower boundary of Tweet decay, we also deployed a second model with a constant $C=0$. Interestingly, the model with a variable $C$ clearly creates a better fit by explaining an additional 8\% of the variance, while having lower AIC, and even lower BIC, which is more sensitive to additional parameters. Consequently,  the monthly decay can be best fit ($R^2 = 0.984, AIC = 260.4, BIC = 266.7$) with:

\begin{equation}
    f(days)=0.40 \cdot exp( -0.0017 \cdot days)+0.46.
\end{equation}

While a lower boundary for Tweet decay of $C=0.46$ is a surprise, a relatively high $C$ is not implausible. We have to keep in mind that at the timescale and time periods we study, we are only capturing some of the processes that can delete a tweet, while systemic events, e.g., a mass exodus of Twitter users to other new platforms, cannot be covered. 

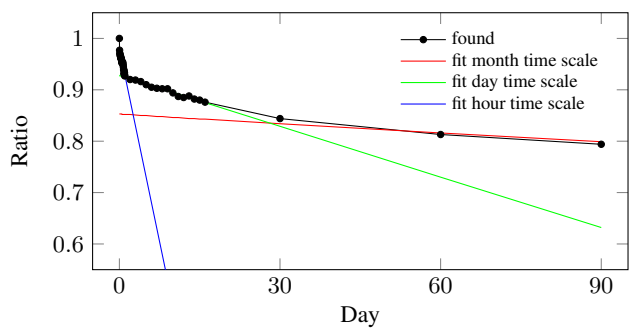
\begin{figure}[b]
\centering
\footnotesize
\pgfplotsset{scaled y ticks=false}
\begin{tikzpicture}
\begin{axis}[
	xlabel=Day,
	ylabel=Ratio,
	width=8.7cm,height=5cm,	xmax=95,ymax=1.05,ymin=0.55,xmin=-5,
	ytick={0.6,0.7,0.8,0.9,1},
	xtick={0,30,60,90},
    y label style={yshift=-.8em}, x label style={yshift=.5em}, 
    legend style={cells={anchor=west},
        draw=none, fill=none, font=\scriptsize,
        legend pos=north east, row sep=-2pt}
    ] 
\addplot [mark=*, mark options={scale=0.6}]
table[x index=0, y index=1] {
0.000 1.000
0.042 0.977
0.083 0.971
0.125 0.969
0.167 0.966
0.208 0.966
0.250 0.962
0.292 0.961
0.333 0.962
0.375 0.960
0.417 0.957
0.458 0.953
0.500 0.952
0.542 0.953
0.583 0.953
0.625 0.954
0.667 0.951
0.708 0.948
0.750 0.944
0.792 0.939
0.833 0.931
0.875 0.938
0.917 0.934
0.958 0.933
1.000 0.928
1.000 0.927
2.000 0.920
3.000 0.919
4.000 0.916
5.000 0.910
6.000 0.905
7.000 0.903
8.000 0.902
9.000 0.902
10.000 0.894
11.000 0.887
12.000 0.885
13.000 0.888
14.000 0.882
15.000 0.880
16.000 0.876
30.000 0.844
60.000 0.813
90.000 0.794
};
\addplot [red]
table[x index=0, y index=1] {
0.000 0.853
0.042 0.853
0.083 0.853
0.125 0.853
0.167 0.853
0.208 0.853
0.250 0.853
0.292 0.853
0.333 0.853
0.375 0.853
0.417 0.853
0.458 0.853
0.500 0.852
0.542 0.852
0.583 0.852
0.625 0.852
0.667 0.852
0.708 0.852
0.750 0.852
0.792 0.852
0.833 0.852
0.875 0.852
0.917 0.852
0.958 0.852
1.000 0.852
1.000 0.852
2.000 0.852
3.000 0.851
4.000 0.850
5.000 0.850
6.000 0.849
7.000 0.848
8.000 0.848
9.000 0.847
10.000 0.846
11.000 0.846
12.000 0.845
13.000 0.844
14.000 0.844
15.000 0.843
16.000 0.843
30.000 0.834
60.000 0.816
90.000 0.799
};
\addplot [green]
table[x index=0, y index=1] {
0.000 0.928
0.042 0.928
0.083 0.927
0.125 0.927
0.167 0.927
0.208 0.927
0.250 0.927
0.292 0.927
0.333 0.927
0.375 0.926
0.417 0.926
0.458 0.926
0.500 0.926
0.542 0.926
0.583 0.926
0.625 0.926
0.667 0.926
0.708 0.925
0.750 0.925
0.792 0.925
0.833 0.925
0.875 0.925
0.917 0.925
0.958 0.925
1.000 0.924
1.000 0.924
2.000 0.921
3.000 0.918
4.000 0.915
5.000 0.911
6.000 0.908
7.000 0.905
8.000 0.901
9.000 0.898
10.000 0.895
11.000 0.892
12.000 0.888
13.000 0.885
14.000 0.882
15.000 0.878
16.000 0.875
30.000 0.829
60.000 0.730
90.000 0.632
};
\addplot [blue]
table[x index=0, y index=1] {
0.000 0.979
0.042 0.977
0.083 0.975
0.125 0.973
0.167 0.971
0.208 0.969
0.250 0.967
0.292 0.965
0.333 0.963
0.375 0.961
0.417 0.959
0.458 0.957
0.500 0.954
0.542 0.952
0.583 0.950
0.625 0.948
0.667 0.946
0.708 0.944
0.750 0.942
0.792 0.940
0.833 0.938
0.875 0.936
0.917 0.934
0.958 0.932
1.000 0.930
1.000 0.930
2.000 0.880
3.000 0.830
4.000 0.780
5.000 0.730
6.000 0.680
7.000 0.630
8.000 0.580
9.000 0.530
10.000 0.481
11.000 0.431
12.000 0.381
13.000 0.331
14.000 0.281
15.000 0.231
16.000 0.181
30.000 -0.517
60.000 -2.013
90.000 -3.510
};
\addlegendentry{found}
\addlegendentry{fit month time scale}
\addlegendentry{fit day time scale}
\addlegendentry{fit hour time scale}
\end{axis}
\end{tikzpicture}
\caption{Proportion of found tweets and additional fitting of short time (hours, days) data decay.}
\label{fig:historic_vanished_short}
\end{figure}

To better understand the differences between the short-time and long-time tweet inaccessibility, we also looked into the rate at which tweets vanish on the hour and day time scales (Figure \ref{fig:historic_vanished_short}). For the hour- and day-time scales, two linear unweighed functions (one for the first 24 hours and one for the first 15 days) showed to be the best fit. 

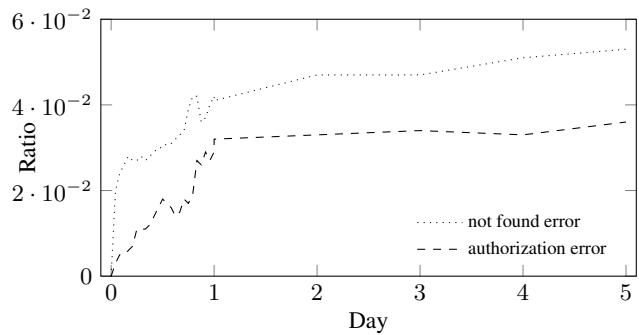
\begin{figure}[t]
\centering
\footnotesize
\pgfplotsset{scaled y ticks=false}
\begin{tikzpicture}
\begin{axis}[
	xlabel=Day,
	ylabel=Ratio,
	width=8.7cm,height=5cm,	xmax=5.1,ymax=0.06,ymin=0.0,xmin=-0.1,
	ytick={0,0.02,0.04,0.06},
	xtick={0,1,2,3,4,5},
    y label style={yshift=-.8em}, x label style={yshift=.5em},
    legend style={cells={anchor=west},
        draw=none, fill=none, font=\scriptsize,
        legend pos=south east, row sep=0pt}
    ]
\addplot [dotted]
table[x index=0, y index=1] {
0.000 0.000
0.042 0.020
0.083 0.024
0.125 0.026
0.167 0.028
0.208 0.027
0.250 0.027
0.292 0.028
0.333 0.027
0.375 0.028
0.417 0.029
0.458 0.030
0.500 0.030
0.542 0.031
0.583 0.031
0.625 0.032
0.667 0.033
0.708 0.034
0.750 0.039
0.792 0.042
0.833 0.042
0.875 0.036
0.917 0.037
0.958 0.040
1.000 0.042
1.000 0.041
2.000 0.047
3.000 0.047
4.000 0.051
5.000 0.053
};
\addplot [dashed]
table[x index=0, y index=1] {
0.000 0.000
0.042 0.003
0.083 0.005
0.125 0.005
0.167 0.006
0.208 0.007
0.250 0.011
0.292 0.011
0.333 0.011
0.375 0.012
0.417 0.014
0.458 0.016
0.500 0.018
0.542 0.017
0.583 0.016
0.625 0.014
0.667 0.015
0.708 0.018
0.750 0.017
0.792 0.019
0.833 0.027
0.875 0.026
0.917 0.029
0.958 0.027
1.000 0.029
1.000 0.032
2.000 0.033
3.000 0.034
4.000 0.033
5.000 0.036
};
\addlegendentry{not found error}
\addlegendentry{authorization error}
\end{axis}
\end{tikzpicture}
\caption{\emph{Not found error} and \emph{authorization error} per hour within the first 24 hours and per day within the first five days.}
\label{fig:historic_vanished_errors}
\end{figure}

Lastly, we investigated what drives the short-time scale tweet inaccessibility (see Fig. \ref{fig:historic_vanished_errors}). A large number of Tweets (2--3\%) vanish within the first hour by deletions (i.e. "not found" error). While we assume that many Tweets are deleted immediately by the author because of typos or on second thought, we also hypothesize that Twitter's recently improved efforts to limit toxicity on the platform may also contribute to this large number of quickly removed Tweets. Interestingly, the proportions of missing Tweets that have been deleted ($\bar{x}=56.3\%, \sigma=0.011$) and that have been protected ($\bar{x}=43.7\%, \sigma=0.011$) are stable over time.

\subsection{Tweet data growth?}
As we could show above, one out of $16$ sent Tweets is not available anymore after $24$ hours. Surprisingly, $3.5\%$ of all Tweets from $day_1$ do not occur in the $day_0$ data. In other words, $1$ out of $~30$ Tweets that we collect $24$ hours after the Tweets where sent, were not available on the Academic API 10 seconds after they were sent. From our knowledge about Twitter's system, this can have two causes:
\begin{enumerate}
    \item Private Tweets are created and then made publicly visible at a later time.
    \item This is an unintended side effect from the technical infrastructure of Twitter.
\end{enumerate}

Because of the high number of Tweets that are affected here, we can conclude that option $1$ is very unlikely and chose to look closer at option $2$. To assess the possibility of technical artifacts that lead to our missing data observations, we extracted some metadata from the Tweet IDs. Kergel et al. \citeyearpar{Kergl2014} described in great detail the process of creating Tweet IDs. A Linux timestamp with milliseconds (with an offset starting with the first Tweet in 2006), a server ID, a worker ID (a server hosts multiple workers), and a sequence number for elements (created at the same millisecond from the same worker on the same server) are combined into a 64 bit number that then gets converted into the known Tweet ID form. This allows us to decode every Tweet ID back into these four variables. The milliseconds of the Tweet are not provided in the \emph{created\_at} variable when collecting Tweets with the API. For the following analysis we compared 108,627 Tweets from $day_0$ with 3,933 Tweets from $day_1$ that were not in the $day_0$ Tweet ID list and we follow the hypothesis that there is a technical reason that causes certain Tweets to arrive later at the API server and are, therefore, not collected right after they were created. We studied all aspects of meta-data coded into the Tweet ID.

\textbf{Location.} A cross tabulation of data centers with the two groups of Tweets shows us that there is no significant deviation from the expected numbers. Table \ref{tab:center} shows us that Twitter currently runs three data centers (IDs 10, 11, and 13) and it seems clear that data center number $10$ shows some indications for missing Tweets, even though the standard residuals (third lines) when comparing observed (first lines) and expected Tweets (second lines) stay below $2.0$ for our data. For further conclusions, it might be interesting to know where these data centers are located. When we look at the Tweet volume over the course of the UTC day for 
data center $10$: \raisebox{-1pt}{\includegraphics[height=2ex,width=12ex]{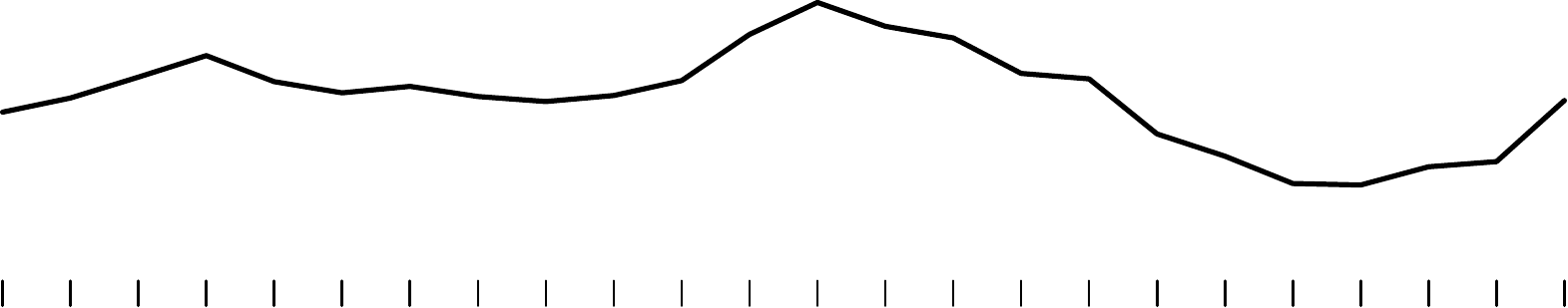}},
data center $11$: \raisebox{-1pt}{\includegraphics[height=2ex,width=12ex]{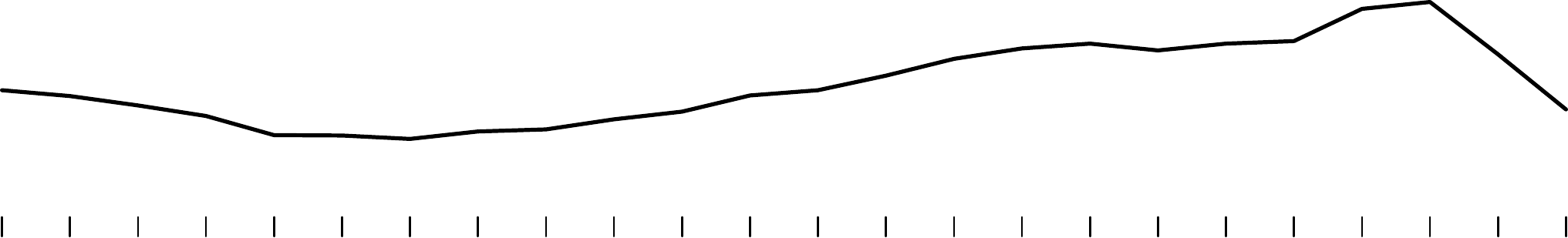}}, and
data center $13$: \raisebox{-1pt}{\includegraphics[height=2ex,width=12ex]{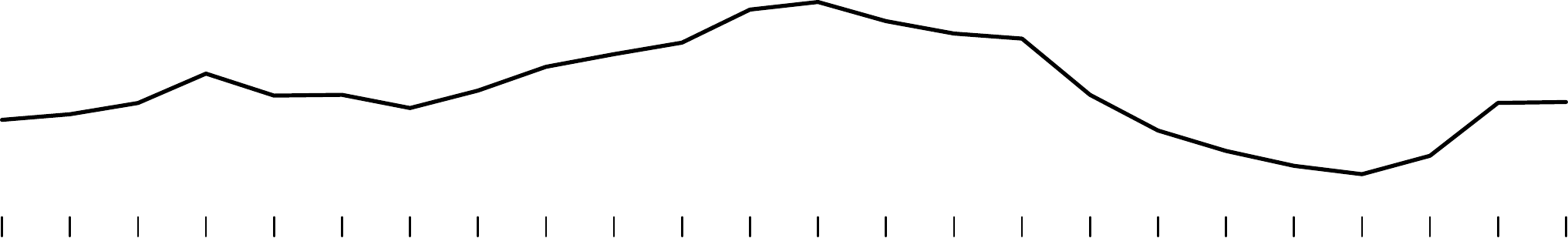}},
we can assume that individual data centers serve geographically different parts of the world, with data center number $11$ most likely serving the United States of America and countries in similar time zones. Interestingly, this would confirm Morstatter et al.'s \citeyearpar{Morstatter2013} observation that Tweets from North America could be over-represented and Tweets from Asia under-represented in Twitter's Streaming APIs.

\begin{table}[t]
\centering
\caption{Cross tabulation of missing Tweets with data center IDs extracted from Tweet IDs: observed and expected number of Tweets, standard residuals.}
\label{tab:center}
\begin{tabular}{lrrrr}
\toprule
 & Center & & & \\
Missing & 10 & 11 & 13 & Total\\
\midrule
No  & 27,235 & 53,332 & 28,060 & 108,627\\
    & 27,294 & 53,284 & 28,049\\
    & -0.356 & +0.209 & +0.063\\
\midrule
Yes &  1,047 &  1,881 & 1,005 & 3,933\\
    &    988 &  1,929 & 1,016\\
    & +1.870 & -1.098 & -0.33\\
\midrule
\textbf{Total} & \textbf{28,282} & \textbf{55,213} & \textbf{29,065} & \textbf{112,560}\\ 
\bottomrule
\end{tabular}
\end{table}

\textbf{Load.} A reason for data delay that leads to missing Tweets in our $day_0$ dataset could be an overburdened worker/server. If this is the case, then Tweets with higher sequence numbers should have a higher chance to be delayed. Or, the other way round, Tweets that are delayed should have on average higher sequence numbers. However, a t-test ($p=0.342$) clearly leads to the rejection of the assumption that the differences in mean are in any way significant. 

\begin{figure}[h]
\centering
  \includegraphics[width=1.0\linewidth]{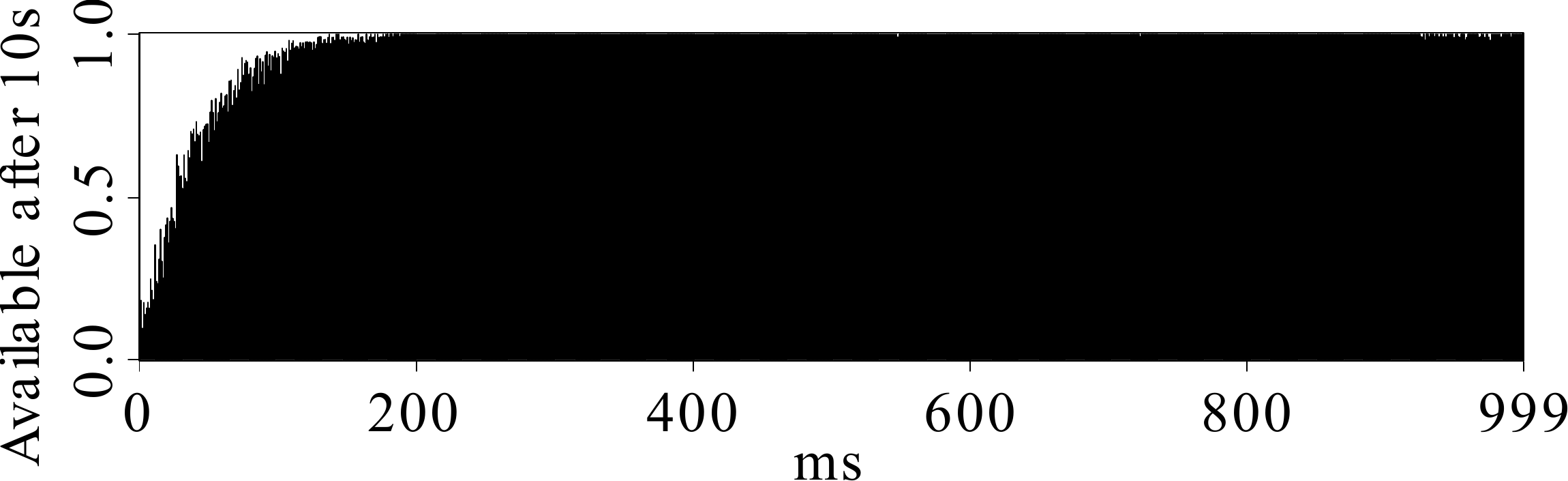}
\caption{Proportion of Tweets per millisecond of timestamp available after $10$ seconds.}
\label{fig:missing} 
\end{figure}

\textbf{Time.} We also took a closer look at the timestamps of the $day_0$ missing Tweets and made a very astonishing observation. While the mean of the millisecond information of all Tweets is, at 502.1, very close to the expected value, the mean millisecond value of the missing Tweets is 40.9---in other words, the majority of Tweets that are missing arrived during the early part of the observed seconds. Figure \ref{fig:missing} shows the extent of this effect. The black area marks the proportion of Tweets per millisecond that are part of $day_0$---for the first couple of milliseconds, ~80\% of Tweets are missing.

Looking at Fig. \ref{fig:missing} might reveal some interesting insights into Twitter's system architecture. If we want to use the Academic API to collect Tweets too close to the current time, we will get an error message from the API that "the time window of data collection must end a minimum of 10 seconds prior to the request time." Consequently, there is a good chance that this API limitation comes from a limitation of the Tweet archival architecture and that the Tweets take some seconds to arrive completely and get indexed at the server/database which handles the Academic API requests---the phenomenon seen in Fig. \ref{fig:missing} could be the result of a stack process in which Tweets get pushed to an API server every second following a last-in first-out procedure. If our observation actually results from such a technical side-effect, we would recommend Twitter to increase the time blocking window in the API to 15 or 20 seconds. 

\section{Discussion}
In the final section, we will discuss some implications for applying the Academic API for research studies. Based on our observations and insight, we also offer some recommendations for working with the Academic API in scientific research projects and for subsequent publishing.

\textbf{New data---new research questions.}
For many researchers, having the restriction to collect only 10 million Tweets per month (the limit at the time of publication of this article) will be perceived as a clear limitation for their data collection efforts. In comparison, we can collect at least the ten-fold number of Tweets with the $1\%$ Sample API. In contrast, the Academic API allows the study of very specific research questions, because we can submit complex queries for a specific time period back to the first Tweets in 2006. This will open Twitter research to new audiences with elevated needs for data quality, especially in the social sciences. At the same time, we will need a growing awareness about what we are actually analyzing when we collect historic Tweets. Twitter's policy of removing all public data from accounts that have been closed by the user or banned by the platform will impact certain research questions. For instance, all these Tweets from Donald Trump that attracted so much global attention over the last years are not on Twitter anymore because the account was suspended from Twitter. And with the removal of the Tweets, all Retweets of these Tweets have been removed as well.

Another aspect that needs to be considered are Twitter's efforts to reducing toxicity on the platform, e.g. Twitter's collaboration with jigsaw to deploy a \emph{Harassment Manager}---a "technology to help women journalists document and manage online abuse\footnote{https://medium.com/jigsaw/technology-to-help-women-journalists-document-and-manage-online-abuse-5edcac127872}". If you want to use Twitter data in the future to study online abuse, you might end up studying the effectiveness of Twitter's efforts to identify and remove this content instead.

\textbf{Data decay.} We have shown in this article that, while the Academic API delivers very good samples, there is a significant drop of available Tweets over time. Unfortunately, this introduces a temporal bias into our data collection, i.e. time periods that are further in the past are under-represented in the dataset. Collecting parts of the data over time could be a possible approach to reduce temporal biasing. In any case, it is clear that this side effect of non-live data collection is hard to overcome and researchers need to be aware of the possible impact on their analyses.

\textbf{Twitter meta data.} When historic Tweets are collected, it is important to realize that most meta data that come with the Tweet or the account that created the Tweet are in the state of the data collection time and not the Tweet creation time. As Zubiaga \citeyearpar{zubiaga_longitudinal_2018} has stated, "The representivity of the metadata, however, keeps fading over time, both because the dataset shrinks and because certain metadata, such as the users' number of followers, keeps changing". In other words, the creation timestamp, the Tweet text, the Tweet ID, and the account ID of the sender stay constant independently from the time of data collection. Other data like the profile information of an account, the information about followers and followees, and even the account name (which can be changed on Twitter) will be from the moment of data collection. This renders questions related to follower networks impossible after a certain time period. On the plus side of getting a sample of the current state of Twitter is that we will find how often every Tweet has been retweeted, replied to, or liked in the meta-data of the Tweet without collecting the Retweets.

\textbf{Open research.} 
It has become a common practice for Twitter research to share the IDs of the Tweets that were used in the analysis of a publication. This practice is also supported by Twitter's terms of use and allows reproducibility to a certain extent.
When using the Academic API, researchers should also document and publish (either in the article or in supplementary material) the following information: Date(s) of data collection, the exact search query, and the utilized API and data endpoint. 
    
\textbf{Content compliance.}
Twitter's developer agreement includes a section called "Content compliance" that says: "If you store Twitter Content offline, you must keep it up to date with the current state of that content on Twitter. Specifically, you must delete or modify any content you have if it is deleted or modified on Twitter." This is very explicit and if you are planning to keep to that agreement, collecting Tweets with the Academic API and then immediately analyzing them is by far the easiest way of doing so.

\textbf{Data minimization.} 
With the API v2 data endpoint, most data fields need to be requested explicitly via call parameters. Intuitively, academic researchers tend to request all possible data fields to maximize the collected dataset. However, it might be worthwhile reconsidering one's data collection strategy with the API v2 endpoint. For instance, when conforming to the EU General Data Protection Regulation (GDPR), personal data shall be "limited to what is necessary in relation to the purposes for which they are processed (‘data minimization’)" [GDPR, Article 5 1.(c)]. Even though Twitter data is often not interpreted as being subject to ethics committees, the Academic API gives us the opportunity to explicitly \emph{not} collect user profile information---especially when this data is irrelevant for our analysis. Furthermore, collecting less data will make the collection process faster.

\textbf{Towards a more collaborative future?}
Twitter has become an important part of political discussions and global dynamics, and scholars in academia use Twitter data for their research. And other scholars cite this research and build on it. By providing data to an enormous number of researchers, Twitter also has to take on responsibility for the quality of the data. This responsibility can be shared with some level of transparency. We are aware that hiding details about algorithms as well as about the software and hardware architecture is a security measure of platform providers. However, we assume that many flaws and curiosities that researchers have found over the last years---e.g. possible regional biases in Twitter's streaming data \cite{Morstatter2013}---could be easily interpreted if we had some level of general information about the system architecture. For instance, at this point, we are confident that Twitter's API architecture is separated from their \emph{regular} system architecture. Consequently, the idea that Tweets from certain areas in the world may be delayed for real-time data collection would be a logical conclusion. Furthermore, some technical artifacts discussed in this or previous articles \cite{Pfeffer2018Tampering} show an indication of not being intended by the system architects. Being more transparent about their systems could lead to researchers identifying bugs and unintended features earlier---which would in return help the platform providers to improve their services.

Our future work will continue helping researchers in assessing the quality of these data sources and will provide further hints for Twitter about how to improve their services. With the release of the Academic API, the improved data endpoint v2, and a more active engagement with academic scholars, Twitter is on the right track to becoming a more trustworthy partner and a more reliable data source for academia.

\section{Ethics Statement}
Our research was not subject to the Ethics Commission at the Technical University of Munich, where the experiments were conducted. Only public Twitter data were used and the publicly available APIs were used. We were not interested in Personal Identifiable Information (PII) in this study and none of the analyses was performed on the individual account level. For those experiments in which we have sent Tweets, we only created Tweets with a single random 42-character string to minimize the risk that our Tweets will manipulate the data collection of other researchers. 

With this study, we have violated Twitter's \emph{Terms of Service}, in particular, the following section: "You may not do any of the following [...] (ii) probe, scan, or test the vulnerability of any system or network ..." However, we strongly believe that our work, as well as similar work, is important to better understand the data that are used by thousands of researchers and practitioners. More transparency about the data will lead to higher quality and more trustworthy research, which in turn, will also be an added value for Twitter. 

\section{Acknowledgements}
We are grateful to Kenny Joseph and Fred Morstatter for their valuable feedback. The lists of Tweet IDs of the experiments of this article can be found here: \url{http://www.pfeffer.at/data/academicapi/}.

\section{Author Contribution}
J.P. planned and carried out the data collection experiments (except Premium API and long-term decay). J.P. and A.M. wrote the manuscript. L.H. contributed important insights into collecting Twitter data, performed the Premium API data collection and provided feedback to the final manuscript. D.G., J.L., and O.S. performed the long-term decay experiment. All authors provided feedback in the manuscript writing process.

\bibliography{aaai23}

\end{document}